\documentclass[aps,prb,twocolumn,showpacs,superscriptaddress]{revtex4-1}
\usepackage{graphicx}
\usepackage{amsmath}

\begin{document}

\title{Low-temperature magnetization in geometrically frustrated Tb$_2$Ti$_2$O$_7$}

\author{E. Lhotel}
\email[]{elsa.lhotel@grenoble.cnrs.fr}
\affiliation{Institut N\'eel, CNRS \& Universit\'e Joseph Fourier, BP 166, 38042 Grenoble Cedex 9, France}
\author{C. Paulsen} \affiliation{Institut N\'eel, CNRS \& Universit\'e Joseph Fourier, BP 166, 38042 Grenoble Cedex 9, France}
\author{P. Dalmas de R\'eotier} \affiliation{Institut Nanosciences et Cryog\'enie, SPSMS, CEA and University Joseph Fourier, F-38054 Grenoble, France}
\author{A. Yaouanc} \affiliation{Institut Nanosciences et Cryog\'enie, SPSMS, CEA and University Joseph Fourier, F-38054 Grenoble, France}
\author{C. Marin} \affiliation{Institut Nanosciences et Cryog\'enie, SPSMS, CEA and University Joseph Fourier, F-38054 Grenoble, France}
\author{S. Vanishri} \affiliation{Institut Nanosciences et Cryog\'enie, SPSMS, CEA and University Joseph Fourier, F-38054 Grenoble, France}

\begin{abstract}
The nature of the low temperature ground state of the pyrochlore compound Tb$_2$Ti$_2$O$_7$ remains a puzzling issue.
Dynamic fluctuations and short-range correlations persist down to 50 mK, as evidenced by microscopic probes. In parallel, magnetization measurements show irreversibilities and glassy behavior below 200 mK.  We have performed magnetization and AC susceptibility measurements on four single crystals down to 57 mK.  We did not observe a clear plateau in the magnetization as a function of field along the [111] direction, as suggested by the quantum spin ice model. In addition to a freezing around 200 mK, slow dynamics are observed in the AC susceptibility up to 4 K. The overall frequency dependence cannot be described by a canonical spin-glass behavior.
\end{abstract}

\maketitle
Geometrical frustration studies in magnetism have revealed a large variety of new and exotic magnetic phases \cite{Lacroix}. Among the frustrated materials, the rare-earth pyrochlore oxides, in which the magnetic atoms are situated on a corner-sharing tetrahedra lattice, form a family in which different kinds of anisotropy and / or magnetic exchange can be obtained, thus providing a playground for studying the role of these parameters in the frustration \cite{Gardner10}. Their chemical formula is R$_2$M$_2$O$_7$ where R is a rare-earth and M a non magnetic metallic element. Among these compounds, Tb$_2$Ti$_2$O$_7$ has been the focus of much interest since no magnetic ordering is observed down to 50 mK \cite{Gardner99, Gardner03} in spite of antiferromagnetic interactions leading to a Curie-Weiss temperature of -13 K \cite{Gingras00}. In this compound, the magnetic Tb$^{3+}$ ions have a strong Ising-like anisotropy along the local [111] axis. Instead of a long range order predicted to occur around 1 K \cite{denHertog00, Kao03}, short range correlations are observed down to the lowest temperature \cite{Gardner99, Fennell12}. 

It has been proposed that the ground state and first excited state of the Tb$^{3+}$ ion are two doublets, separated by a gap $\Delta \simeq 18$ K \cite{Gingras00, Mirebeau07}. In this framework, it was suggested that quantum induced crystal field excitations could help explain the absence of long-range order in Tb$_2$Ti$_2$O$_7$ \cite{Molavian07}. Thus this compound was considered to be a "quantum spin ice" in analogy to dipolar spin ices which are characterized by a degenerate ground state in which two spins point in and two spins point out of a tetrahedron \cite{Bramwell01}. It follows that the magnetization curves at low enough temperatures should show a magnetization plateau in the presence of an applied field along the [111] direction \cite{Molavian09} as observed in dipolar spin ice compounds \cite{Sakakibara03}. Notwithstanding, recently it has been proposed that a tetragonal distortion which occurs at low temperature \cite{Mamsurova86, Ruff07, Lummen08} lifts the degeneracy of the lowest doublet, thus generating a singlet ground state \cite{Rule09, Chapuis09}. In this case, the resulting Tb magnetic moment is induced by exchange interactions \cite{Bonville11}. 
As the nature of the Tb$^{3+}$ ion ground state is still a debated question \cite{Gaulin11}, magnetization curves at very low temperature could be a key to resolving it. 

Another puzzling feature of the Tb$_2$Ti$_2$O$_7$ low temperature magnetic state is the apparent contradiction between the persistent dynamics down to 50 mK \cite{Gardner99}, and the evidence of  freezing observed near 200 mK by AC susceptibility \cite{Gardner03, Hamaguchi04} and in Zero Field Cooled-Field Cooled (ZFC-FC) magnetization measurements \cite{Luo01}, which was at first attributed to the presence of defects \cite{Gardner03}. Irreversibilities in muon spin relaxation ($\mu$SR) measurements and a minimum in specific heat have been reported at the same temperature \cite{Yaouanc11}. Furthermore, a saturation of the muon relaxation rate has been observed below about 2 K \cite{Gardner99, Yaouanc11}. These observations suggest the coexistence of different time and temperature scales for the spin dynamics, and up to now, no clear unified picture has emerged to describe the  dynamics of Tb$_2$Ti$_2$O$_7$. Although microscopic probes (neutron scattering and $\mu$SR measurements) have been extensively used, detailed studies of the magnetization \cite{Legl12} and of the AC susceptibility need to be performed at very low temperature.

In this communication we present a systematic study of the macroscopic magnetic properties down to 57 mK made on Tb$_2$Ti$_2$O$_7$ single crystals. We show that down to this temperature, there is no evidence of a magnetization plateau in the [111] direction \cite{Molavian09}. We also show that the reported glassy phase \cite{Gardner03, Hamaguchi04} is not a canonical spin glass phase, and is strongly dependent on the applied field.

Magnetization and AC susceptibility measurements were performed by the extraction method, using two superconducting quantum interference device magnetometers equipped with miniature dilution refrigerators and developed at the Institut N\'eel \cite{Paulsen01}. AC measurements were carried out for frequencies between 1.1~mHz and 2.11 kHz, with an applied AC field of 0.5 and 4 Oe (No dependence on the amplitude of the AC field was observed). 

Three crystals, denoted as B, C, and D, were prepared for these measurements (see Suppl. Mat.). 
These crystals have been characterized by specific heat measurements. The data for crystal C have been published in Ref. \onlinecite{Yaouanc11}. The specific heat behaviors of samples B and C are qualitatively the same, with no anomalies, although sample B has a slightly larger residual entropy. In contrast, the specific heat of sample D is very different with a narrow peak around 400 mK, reminiscent of the data published in Ref. \onlinecite{Hamaguchi04}. These crystals were cut into flat, disk shaped samples of about 30 mg. Four samples were measured: two from crystal C, called C1 and C2 measured along the [111] and [110] directions respectively, one from crystal B, and one from crystal D, both measured along the [111] direction. 
Demagnetizing effects are very small due to the geometry of the samples (the field orientation was in the plane of each disk, except for sample D). 
The crystal alignment with respect to the field is accurate to within a few degrees. 

\begin{figure}
\includegraphics[keepaspectratio=true, width=8cm]{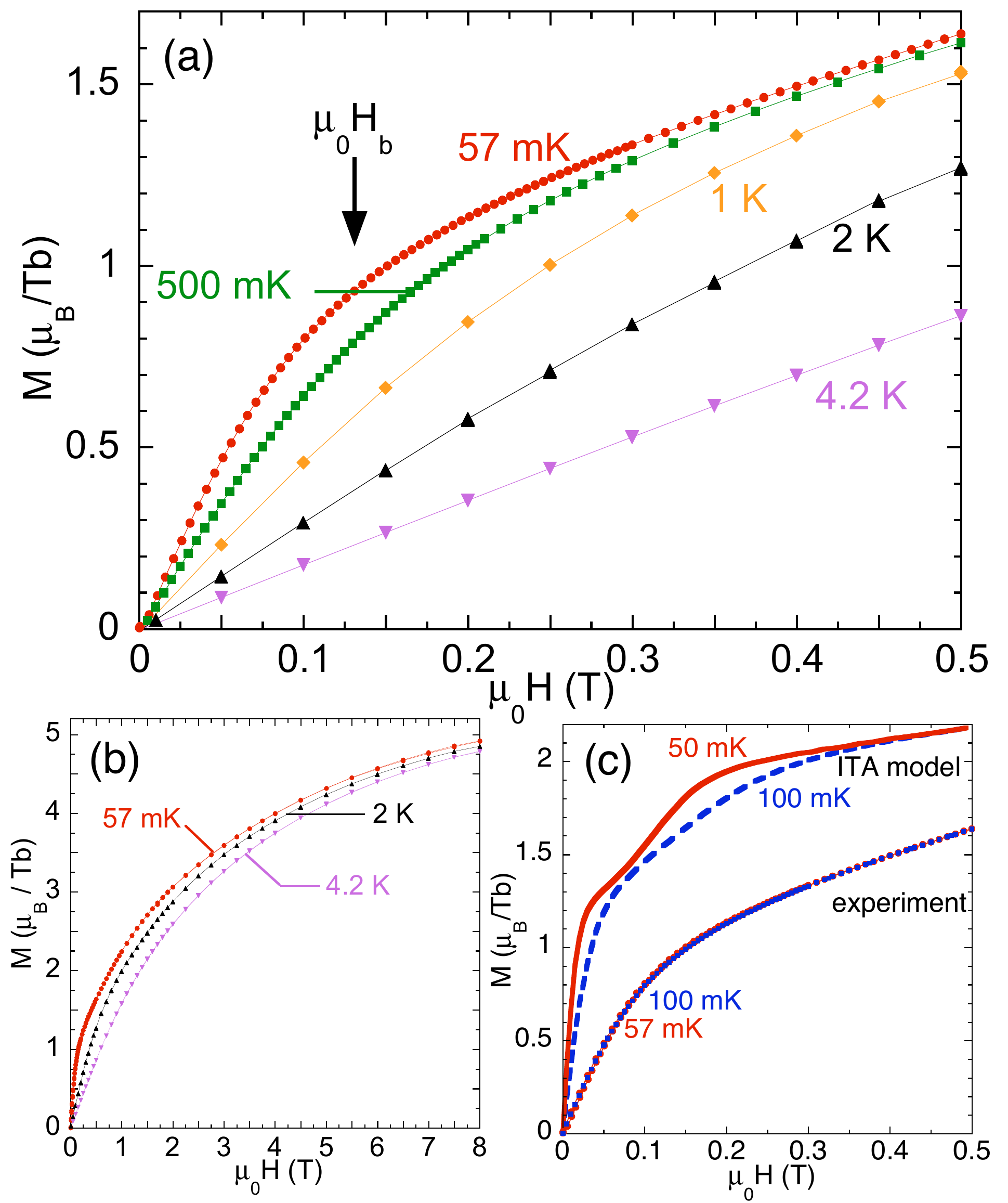}
\caption{(Color online) First magnetization curves $M$ vs. $\mu_0 H$ along the [111] direction between 57 mK and 4.2 K for sample B (a) up to 0.5 T, (b) up to 8 T. (c) Comparison between the experimental curves at 57 and 100 mK to those predicted in Ref. \onlinecite{Molavian09}.}
\label{figMH111}
\end{figure}

Measurements of the first magnetization curves along the [111] direction were carried out on samples B, C1 and D. Figure \ref{figMH111}(b) shows these isotherms $M$ vs. $\mu_0 H$ for fields up to 8 T between 57 mK and 4.2 K for sample B. The magnetization reaches 5 $\mu_B$/Tb  at 8 T, which is consistent with previous measurements \cite{Mirebeau07}, and corresponds to the predicted ground state magnetic moment for the Tb ion \cite{Gingras00}. The results are very similar for the three samples (See Suppl. Mat.). 

Figure~\ref{figMH111}(a) shows the magnetization below 0.5 T in the same temperature range. These curves can be compared with the predictions of Molavian and Gingras \cite{Molavian09} who proposed the emergence of a magnetization plateau below 100 mK in the [111] direction. At 57 mK, no clear evidence of such a plateau can be observed, which is confirmed by plotting the calculated derivative $dM/dH$ vs $H$ (See Suppl. Mat.). 
Nevertheless a small break of slope occurs around $\mu_0 H_b=0.13$ T. This field can be compared to the level crossing field $B_c$=0.082 T in Ref. \onlinecite{Molavian09}, which also gives rise to a change of curvature in the calculated 100 mK magnetization curve (See Fig. \ref{figMH111}(c)). 
In addition, we have measured the magnetization curves along the [110] direction down to 70 mK in sample C2: they are only slightly different from the magnetization curves measured along [111] (see Suppl. Mat.). 

Two main differences emerge when we compare our measured curves along [111] to the predicted magnetization curves of Ref. \onlinecite{Molavian09} calculated in the independent tetrahedra approximation (ITA) (See figure \ref{figMH111}(c)): i) The measured value of the magnetization at 0.5 T is significantly lower than the calculated magnetization. ii) More important, our 57 and 100 mK curves are almost identical whereas the calculated 50 and 100 mK curves have very different shapes. 

Several hypotheses can be proposed to account for these discrepancies. i) It has been suggested that the exchange interaction is anisotropic \cite{Sazonov10, Bonville11}. In this scheme the proposed exchange energies are larger than the isotropic exchange constant used in the calculation of Ref. \onlinecite{Molavian09}. Ref. \onlinecite{Molavian09} shows that the increase of the exchange  tends to decrease the magnetization value at low field and to decrease the temperature at which the plateau is formed. Thus in this case, our lowest measuring temperature could be higher than the temperature for the formation of the plateau.  ii) It has been recently proposed that the Tb ground state is a singlet \cite{Chapuis09, Rule09, Bonville11} rather than a doublet. This scenario presumably precludes the existence of the plateau. The computation of the magnetization curves within this crystal field scheme would be of great interest to compare to the present data.
 iii) In Ref. \onlinecite{Molavian09}, the calculation were made in the independent tetrahedra approximation. It would be interesting to check the robustness of the plateau when including further neighbors.  

\begin{figure}
\includegraphics[keepaspectratio=true, width=8cm]{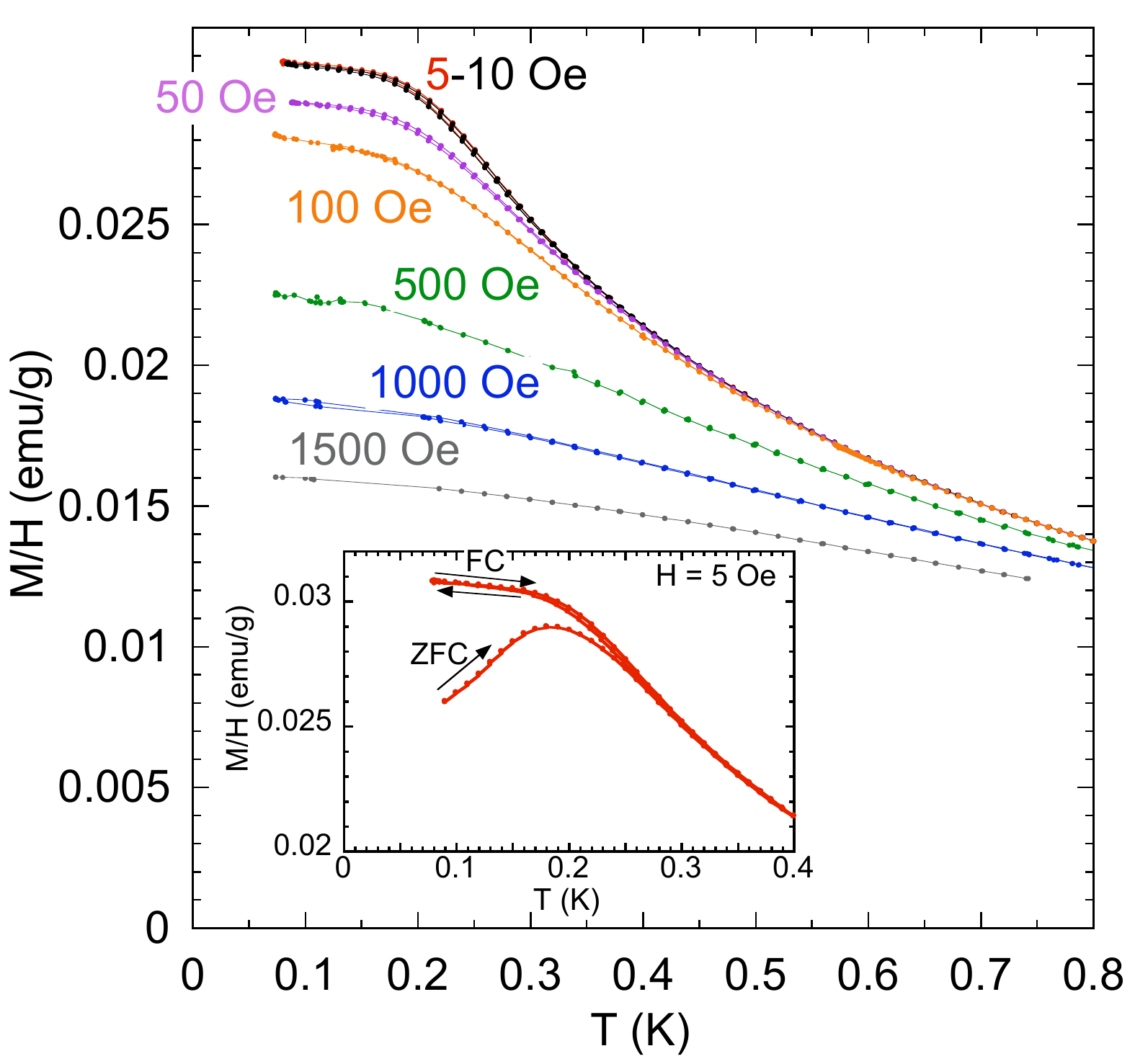}
\caption{(Color online) Field Cooled $M/H$ vs. $T$ along the [110] direction for sample C2 at different applied fields. Inset: $M/H$ vs $T$ with $H=5$ Oe in a ZFC-FC procedure.}
\label{figMT}
\end{figure}

We measured the ZFC-FC magnetization as a function of temperature for the four samples. All of the data show an irreversibility below 250~mK (See inset of Fig. \ref{figMT}):  the ZFC curve has a maximum between 150 and 200 mK, whereas the FC magnetization tends to saturate. This behavior is in agreement with previous magnetization measurements \cite{Luo01}. It occurs at almost the same temperature as the irreversibility observed in transverse $\mu$SR measurements (with a 600 Oe field along [110]) as well as the minimum in specific heat \cite{Yaouanc11}. When increasing the magnetic field above 50 Oe, the magnetization starts to saturate, thus leading to a smaller value of $M/H$ (See Fig. \ref{figMT}), and the irreversibility is reduced. 

\begin{figure}
\includegraphics[keepaspectratio=true, width=7cm]{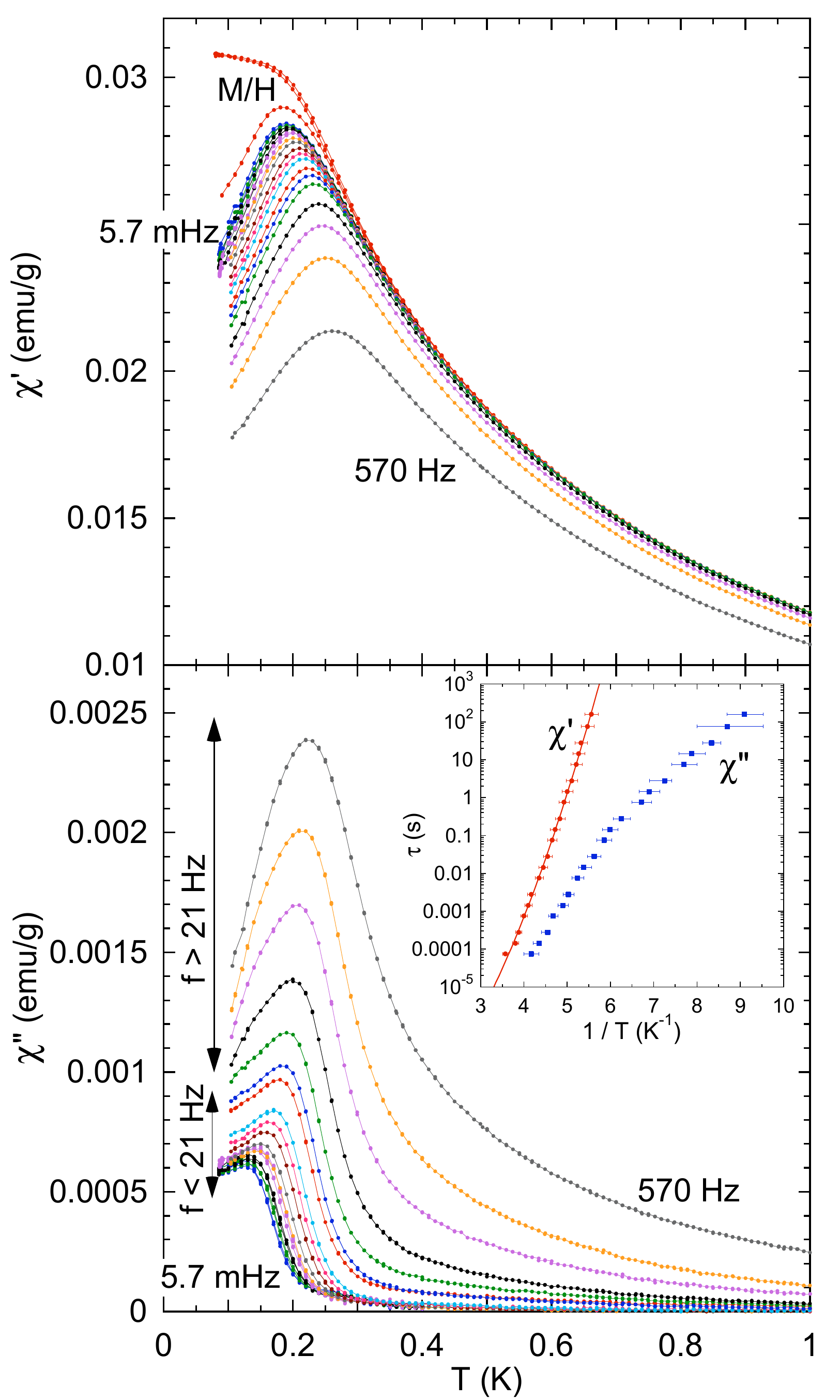}
\caption{(Color online) AC susceptibility measurements along the [110] direction for sample C2. Top: $\chi'$ vs. $T$ together with $M/H$ vs. $T$ in a ZFC-FC procedure with $H_{DC}=5$ Oe. Bottom: $\chi"$ vs. $T$. $\chi_{AC}$ is plotted for frequencies between 5.7 mHz and 570 Hz with $H_{AC}=0.5$ Oe. Inset: $\tau=1/(2\pi f)$ vs. $1/T$, where $T$ is the temperature of the maximum of $\chi'$ (red circles) and of $\chi"$ (blue squares). The line is a fit to the equation $\tau=\tau_0 \exp [(E/T)^{\sigma}]$ with $\tau_0=1.1\times 10^{-9}$ s, $E=0.91$ K and $\sigma=2$. }
\label{figXac}
\end{figure}

Figure \ref{figXac} shows an investigation of the frequency dependence of the AC susceptibility associated with this irreversibility. 
We measured the AC susceptibility along the [111] and [110] directions for samples C1 and C2 respectively, and the behavior is identical for both directions. 
As seen in Figure \ref{figXac}, the real part $\chi'$ of the susceptibility as a function of temperature presents a well-defined peak, which occurs at the same temperature as in previous measurements \cite{Gardner03, Hamaguchi04}. The imaginary (or dissipative) part  $\chi"$ is about 10 \% of $\chi'$, indicating that a substantial amount of the sample is involved in these slow dynamics. 
For samples B et D, the peak in $\chi'$ occurs in the same range of temperature. Nevertheless, it is broader and $\chi''$ exhibits a more complex shape (see Suppl. Mat.). These features indicate that the physical origin of the freezing is intrinsic to the compound and that only the detailed dynamics might be affected by the slight differences between the crystals. 

The AC susceptibility behavior is very different from a canonical spin-glass behavior. Usually, a maximum in $\chi'(T)$ is caused by a competition between the increase of the magnetic polarization and the slowing down of the dynamics when the temperature decreases. The existence of a distribution of relaxation times broadens the maximum and makes it difficult to compare $\chi'$ to usual dynamic laws because the freezing temperature $T$, which can be defined as the $\chi'$ maximum\cite{footnote1}, does not provide a direct determination of the maximum relaxation time $\tau$ \cite{Prejean88}.
Nevertheless, in canonical spin glasses, two relations are often used to describe the frequency dependence of the $\chi'$ maximum  \cite{Souletie85}: i) the Vogel-Fulcher law $\tau = \tau_{0} \exp[E/(T-T_0)]$, where $E$ is an energy barrier, $T_0$ is a phenomenological parameter, $\tau_0$ is the intrinsic relaxation time, and $\tau$ is related to the measurement frequency $\tau=1/(2 \pi f)$, and ii) the dynamic scaling law  $\tau = \tau_0 [(T-T_c)/T_c]^{z \nu}$, where $T_c$ is the transition temperature and $z \nu$ is the dynamic critical exponent.
In the case of Tb$_2$Ti$_2$O$_7$, neither of these laws fits the temperature dependence of the $\chi'$ maximum, thus indicating the absence of a canonical spin-glass transition. 
However, the law $\tau=\tau_0 \exp [(E/T)^\sigma]$ fits the data with $\tau_0 \approx 1.1 \times 10^{-9}$ s, $E \approx 0.91$ K and $\sigma \approx 2$ (See the inset of Fig.~\ref{figXac}). Such a law has been proposed in the context of a zero temperature spin-glass transition \cite{McMillan84, Young83}. In this case, the maximum of the susceptibility is related to a glassy behavior rather than to a spin-glass phase transition. We can note that although this detailed analysis of the frequency dependence of the $\chi'$ peak shows that Tb$_2$Ti$_2$O$_7$ does not exhibit a canonical spin-glass transition, the peak shift per decade frequency $\Delta T/[T\Delta(\log f)]$ is found to be 0.06-0.08 (depending on the frequency range) and is thus similar to that for insulating spin-glasses \cite{Mydosh}. 

In addition to this analysis of $\chi'$, it is interesting to focus on the dissipative part $\chi"$ (See the bottom of Fig.~\ref{figXac}). In a simple thermally activated process with a single energy barrier $E$, the frequency dependence of the maximum of $\chi"$ as a function of $1/T$ would follow an Arrhenius law $\tau=\tau_0 \exp(E/T)$.
This is not the case (See the inset of  Fig. \ref{figXac}), which suggests a more complicated process. The unusual shape of the $\chi''(T)$ curves confirms it. 
 At low frequency, $f<21$~Hz, the $\chi"$ onset occurs at low temperature ($T<0.4$ K) and the peak is quite sharp and asymmetric. At higher frequency, we enter a second regime in which the peak is much more rounded. This might be due to an additional relaxation process at these frequencies. Indeed, an intriguing feature arises in this regime: The value of $\chi"$ decreases very slowly when increasing the temperature and falls to zero at much higher temperature. For example, for frequencies larger than 570 Hz, the $\chi"$ onset occurs at temperatures larger than 4 K. This "high frequency" regime is observed in all the samples and indicates that slow dynamics emerges at a much higher temperature than the effective freezing (200 mK) of the compound. Since it was shown that correlations at 2.5 K extend only over a single tetrahedron \cite{Gardner99}, this high temperature dissipation might be related to the dynamics of these single tetrahedra. 

These two regimes can be related to previous microscopic measurements, although in different time scales.  
When decreasing the temperature from 10 to 1 K, the $\mu$SR asymmetry shows a change in shape and the $\mu$SR relaxation rate $1/T_1$ increases rapidly \cite{Gardner99, Yaouanc11}. Meanwhile, the neutron scattering quasi-elastic energy linewidth decreases \cite{Yasui02, Yaouanc11}. Together with the onset of $\chi"$ around 4 K at "high" frequencies, these results indicate that, in this temperature range, several characteristic times exist in the system (from 10$^{-11}$ to 10$^{-2}$ s) and that all of them are increasing. 
At lower temperatures, below 1 K, the $\mu$SR relaxation rate starts to saturate. In addition, neutron spin echo (NSE) measurements show that, below 400 mK, more than 10 \% of spins are nearly frozen within the NSE time scale window ($< 1$ ns) \cite{Gardner03}. The low temperature peak in AC susceptibility measurements might be induced by these "frozen" spins at the NSE scale. 
These observations show that the original magnetic ground state observed in Tb$_2$Ti$_2$O$_7$ results in a rich dynamic behavior over a very broad time range, from less than the nanosecond to the quasistatic regime. We can wonder whether these dynamics could be induced by the existence of multispin excitations. Indeed, loops with a broad distribution of lengths were identified in the Coulomb phase of spin ice \cite{Jaubert11}. In kagome systems, similar loops were shown to have temperature dependent relaxation times and to induce a dynamical freezing \cite{Cepas12}. 

It is worth noting that the shape of our AC curves, especially the $\chi"$ part, is very similar to recent measurements in Tb$_2$Sn$_{1.8}$Ti$_{0.2}$O$_7$ \cite{Dahlberg11}. It shows that the slow dynamics probed by AC susceptibility is robust against deformation and confirms that is is not simply due to defects or impurities in the samples. 

\begin{figure}[h]
\includegraphics[keepaspectratio=true, width=8cm]{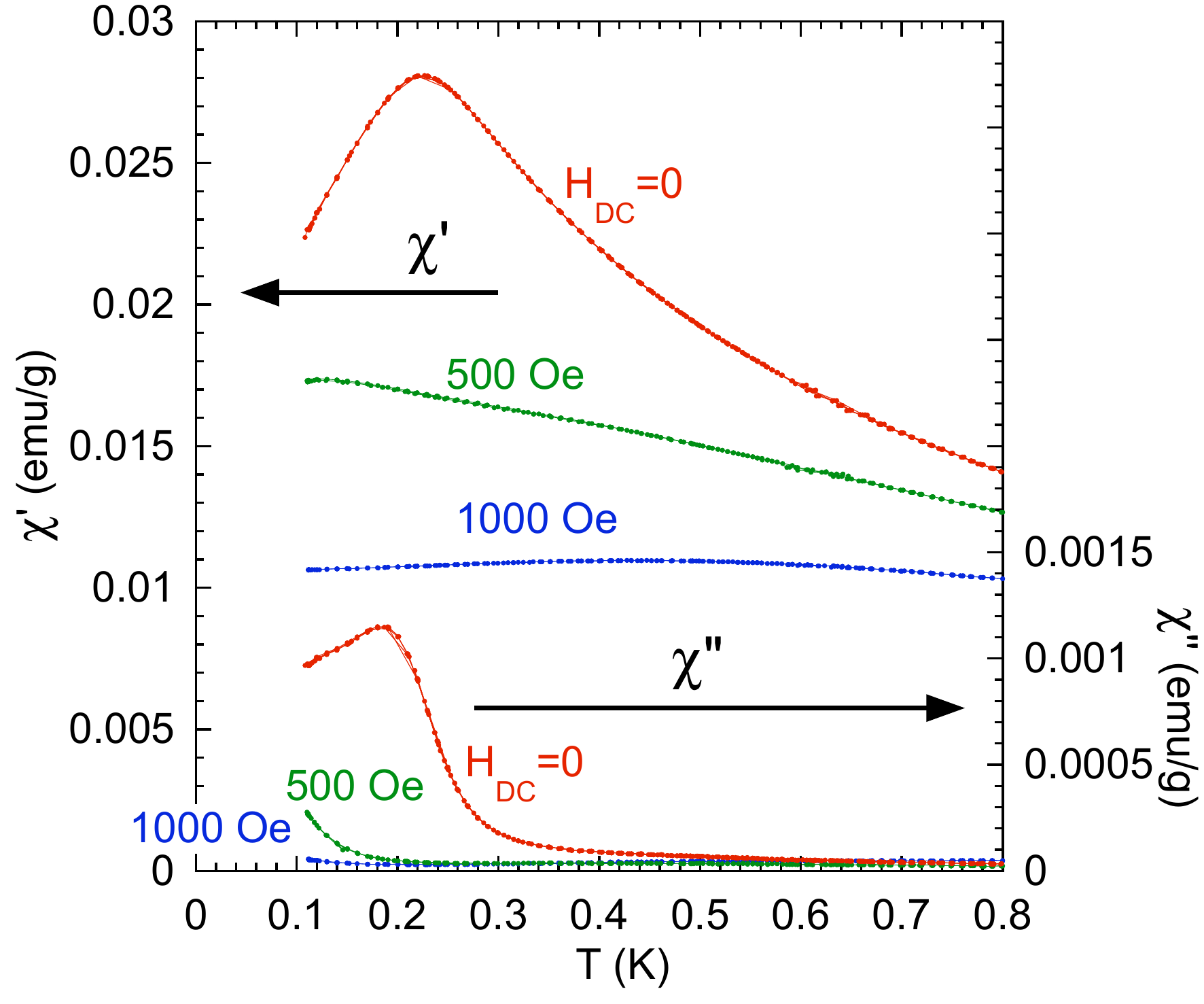}
\caption{(Color online) $\chi'$ (left axis) and $\chi"$ (right axis) vs. $T$ along the [111] direction for sample C1, with $H_{AC}=4$ Oe and $f=11.1$ Hz, in different applied DC fields.}
\label{figXT_H}
\end{figure}

Finally, the AC susceptibility measurements with a DC applied field show a very complex behavior (See Fig. \ref{figXT_H}).  i) The $\chi'$ signal is much smaller than the corresponding $M/H$ amplitude, showing that a part of the spins are easily polarized by the DC field, become blocked and cease to respond to the AC field.  In parallel, the application of a DC field strongly suppresses $\chi"$ at low frequency. 
These results are related to saturation effects and to the suppression of the irreversibility in the ZFC-FC curves reported above when increasing the magnetic field. ii) On the contrary, the high temperature part of $\chi"$ at high frequency ($f>21$ Hz) remains important, even in presence of a 1000 Oe DC field (not shown). The origin of this high frequency broad signal remains unclear, but it might be a key feature in understanding the Tb$_2$Ti$_2$O$_7$ exotic magnetic state, especially since it seems robust against magnetic field. 

In conclusion, we have shown that the magnetization curves of Tb$_2$Ti$_2$O$_7$ measured along the [111] direction do not show evidence of a plateau down to 57 mK, contrary to the predictions of the quantum spin ice model \cite{Molavian09}, and in agreement with Legl {\it et al.} \cite{Legl12}. Moreover, the [110] magnetization curves are very similar to the [111] data. Below 250 mK, the ZFC-FC curves present an irreversibility which is reduced when increasing the applied magnetic field. Our analysis of the frequency dependence of the AC susceptibility leads us to conclude that these dynamics do not correspond to a canonical spin-glass transition, but rather to a freezing. The unusual features observed in the $\chi"$ part of the susceptibility up to 4 K show that the slowing down of the dynamics occur at much higher temperature than the freezing temperature. The compatibility between our measurements on four samples and previous studies make us believe that these features are intrinsic, and that the coexistence of fast and slow dynamics is a clue in understanding the physics of Tb$_2$Ti$_2$O$_7$.

We would like to thank R. Ballou, B. Canals, J.-J. Pr\'ejean and V. Simonet for useful discussions.

\clearpage

{\parbox[c]{17cm}{{\large \bf \begin{center}Supplemental Material\end{center}}

\bigskip
In studies of Tb$_2$Ti$_2$O$_7$, various specific heat responses have been reported in different samples \cite{Chapuis,  Chapuis10_SM, Takatsu12}.  We have measured the magnetization and AC susceptibility of several samples with different specific heat characteristics. This Supplemental Material presents these measurements performed on four samples, called B, C1, C2 and D. No important differences were observed between the samples, thus confirming our conclusions. }}


\setcounter{figure}{0}

\section{Sample preparation}
Polycrystalline Tb$_2$Ti$_2$O$_7$ was first synthesized from Tb$_2$O$_3$ and TiO$_2$ powders (4N and 4N5). 
Sample C was made using commercial Tb$_2$O$_3$. 
For samples B and D, Tb$_2$O$_3$ was obtained by reduction of a commercial powder of Tb$_4$O$_7$ (4N) under a flow of hydrogen and argon gas in the molar ratio 1/9.

For the three crystals an initial heat treatment of the Tb$_2$O$_3$ and TiO$_2$ powders to 1200$^{\circ}$C was followed by a second treatment up to 1350$^{\circ}$C with an intermediate grinding and compaction so as to obtain a dense rod. The crystals were subsequently prepared from the rods of Tb$_2$Ti$_2$O$_7$ using the traveling floating zone technique. Crystals B and D were grown under one atmosphere of argon with growth velocities of 8 and 3 mm/h respectively, and crystal C was prepared under oxygen gas with a velocity of 7 mm/h. 

\begin{figure}[h!]
\includegraphics[keepaspectratio=true, width=8cm]{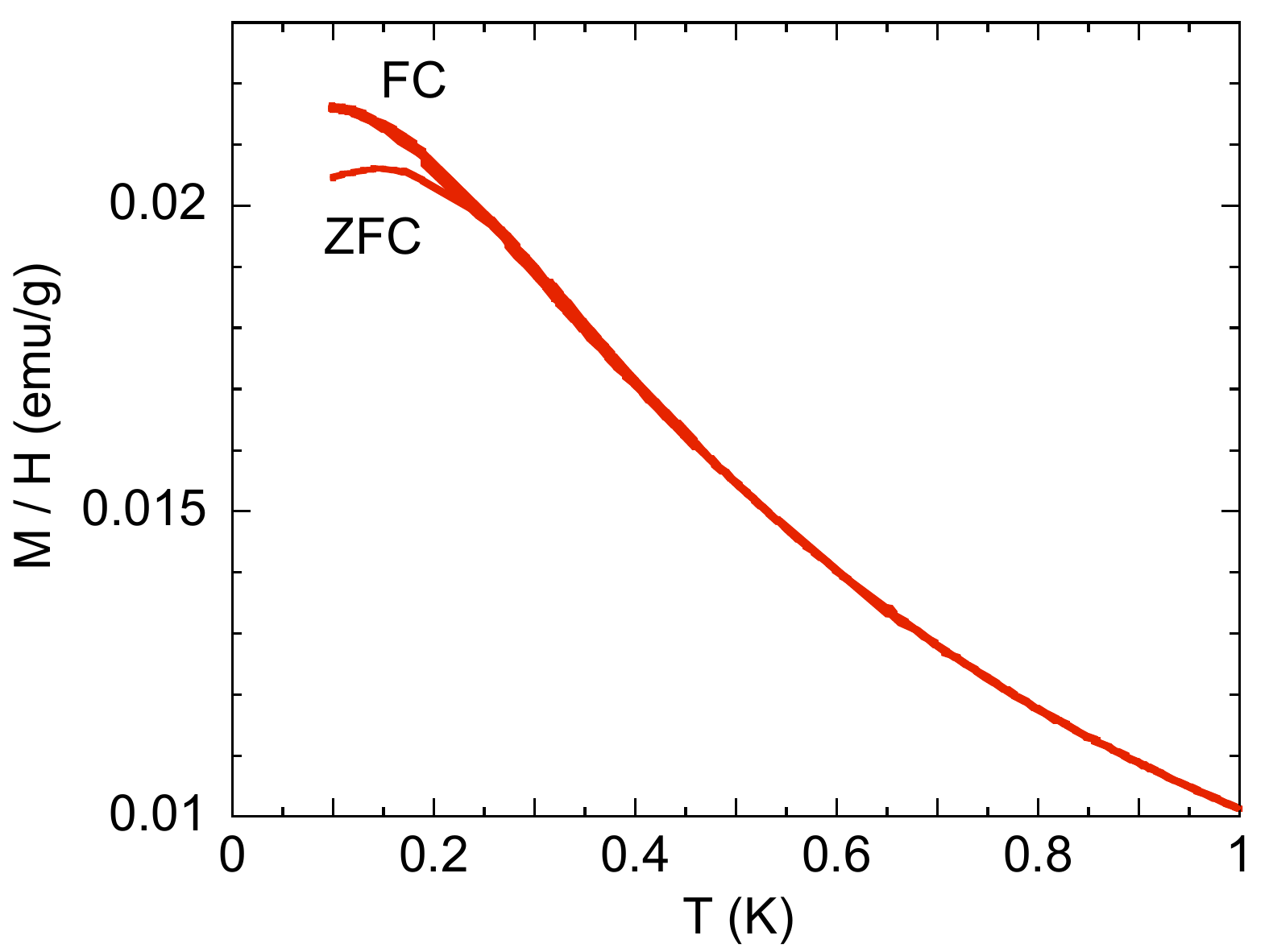}
\caption{(Color online) $M/H$ vs $T$ in ZFC-FC procedure for sample D, measured in an applied field of 100 Oe along [111]. }
\label{figMTD}
\end{figure}

These crystals were characterized by heat capacity measurements \cite{Chapuis, Yaouanc11_SM}. We recall that sample D presents an anomaly in the specific heat around 400 mK, which is not associated with any anomaly in the magnetization at this temperature (See Figure \ref{figMTD}). The specific heat of samples B and C are very similar, but sample B has a slightly higher residual entropy.

\raisebox{-3.1 cm}{}
\section{Magnetization measurements along [111] in different samples}
\begin{figure}[h!]
\includegraphics[keepaspectratio=true, width=7cm]{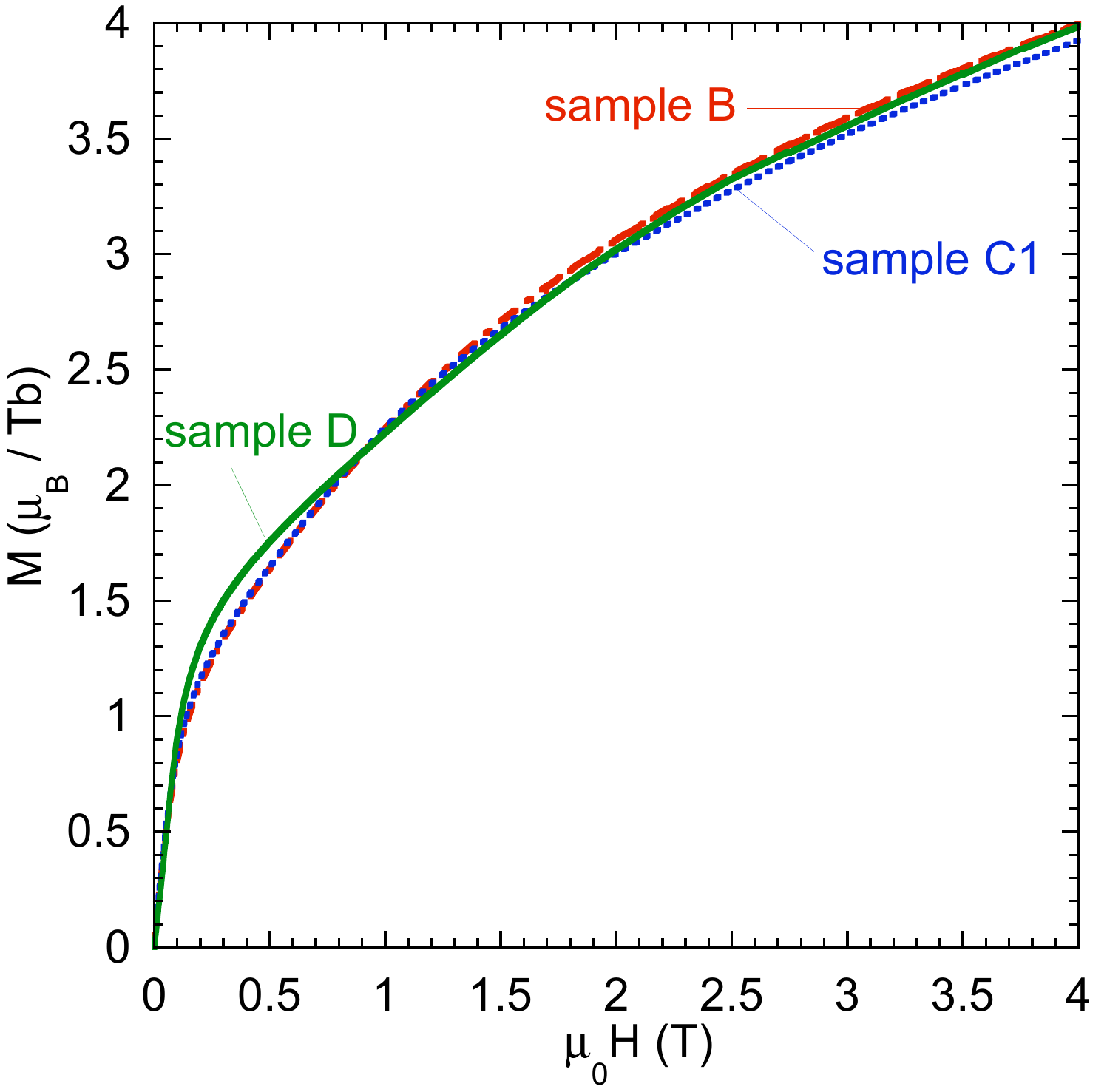}
\caption{(Color online) First magnetization curves $M$ vs. $\mu_0 H$ at 80 mK along the [111] direction. Red dashed line: sample B. Blue dotted line: sample C1. Green line: sample D. }
\label{figMH3samples}
\end{figure}

\begin{figure}[h!]
\includegraphics[keepaspectratio=true, width=7cm]{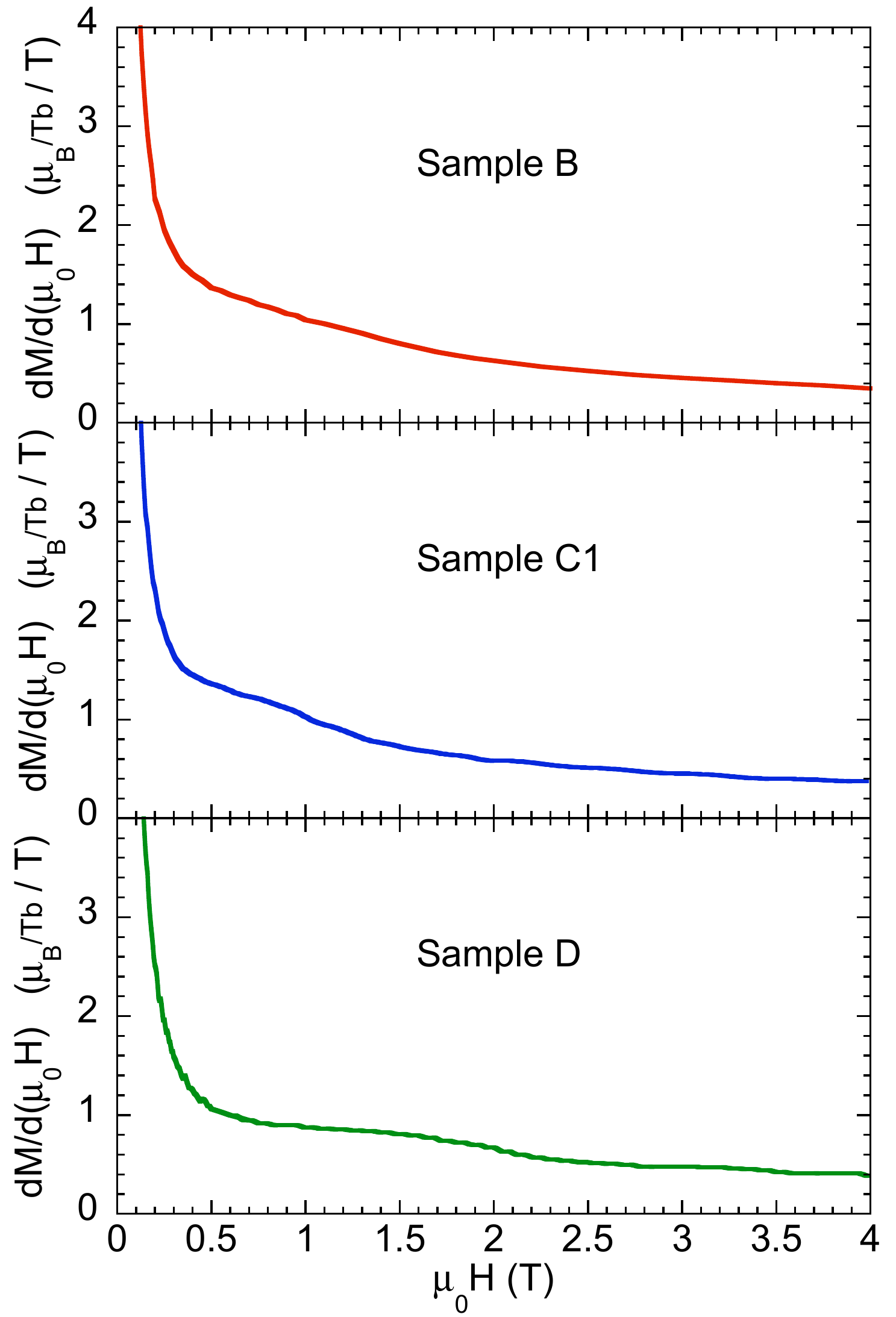}
\caption{(Color online) $dM/dH$ vs. $\mu_0 H$ at 80 mK along the [111] direction in samples B, C1 and D. }
\label{figdMdH3samples}
\end{figure}

Figure \ref{figMH3samples} shows the magnetization curves at very low temperature when the field is applied along the [111] direction for three samples: B, C1 and D. We can see that samples B and C1 present almost identical magnetization curves, whereas sample D shows a slightly different curvature around 0.3 T. To clarify these differences, we have plotted the derivative $dM/dH$ as a function of field in Figure \ref{figdMdH3samples}. 

\section{Magnetization measurements along [110] vs. [111]}

\begin{figure}[h!]
\includegraphics[keepaspectratio=true, width=7cm]{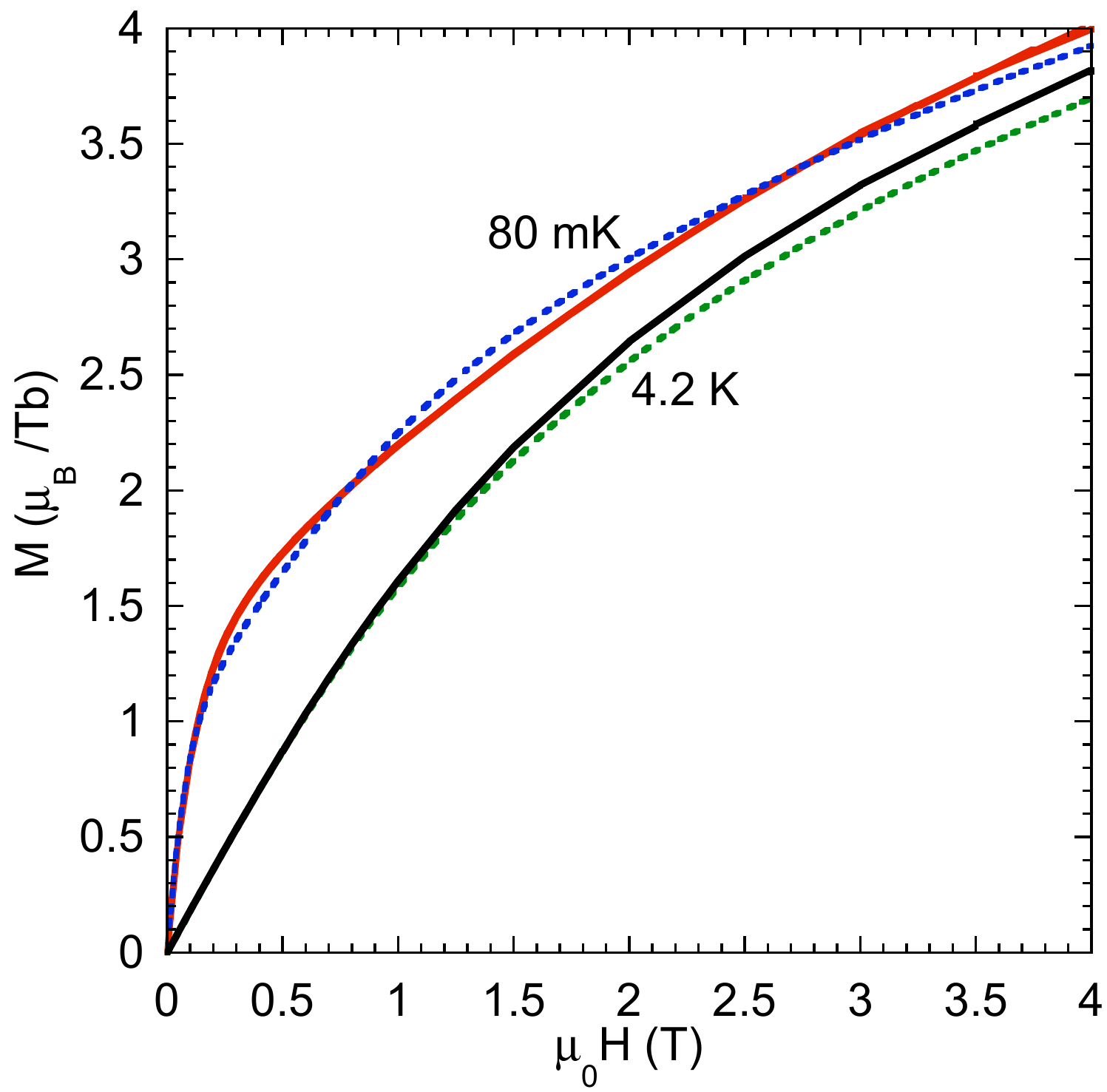}
\caption{(Color online) First magnetization curves $M$ vs. $\mu_0 H$ at 80 mK and 4.2 K. Solid lines: sample C2, measured along the [110] direction. Dotted lines: sample C1, measured along the [111] direction. }
\label{figMHC12}
\end{figure}

Figure \ref{figMHC12} presents the magnetization curves of samples C1 and C2 measured along the [111] and [110] directions respectively (Crystal C is the the crystal with the smallest residual entropy). 

The initial slope is very similar in both directions. 
At low temperatures, the magnetizations  cross three times. Furthermore, the magnetization along the [110] direction is slightly larger than along [111] at 8 T, even at 4.2~K. The latter observation is consistent with other measured magnetization curves \cite{Yasui02_SM, Legl12_SM} but was not reproduced by calculation \cite{Klekovkina11}. 

\vspace{0.6cm}

\section{AC susceptibility in sample B}
\begin{figure}[h!]
\includegraphics[keepaspectratio=true, width=7cm]{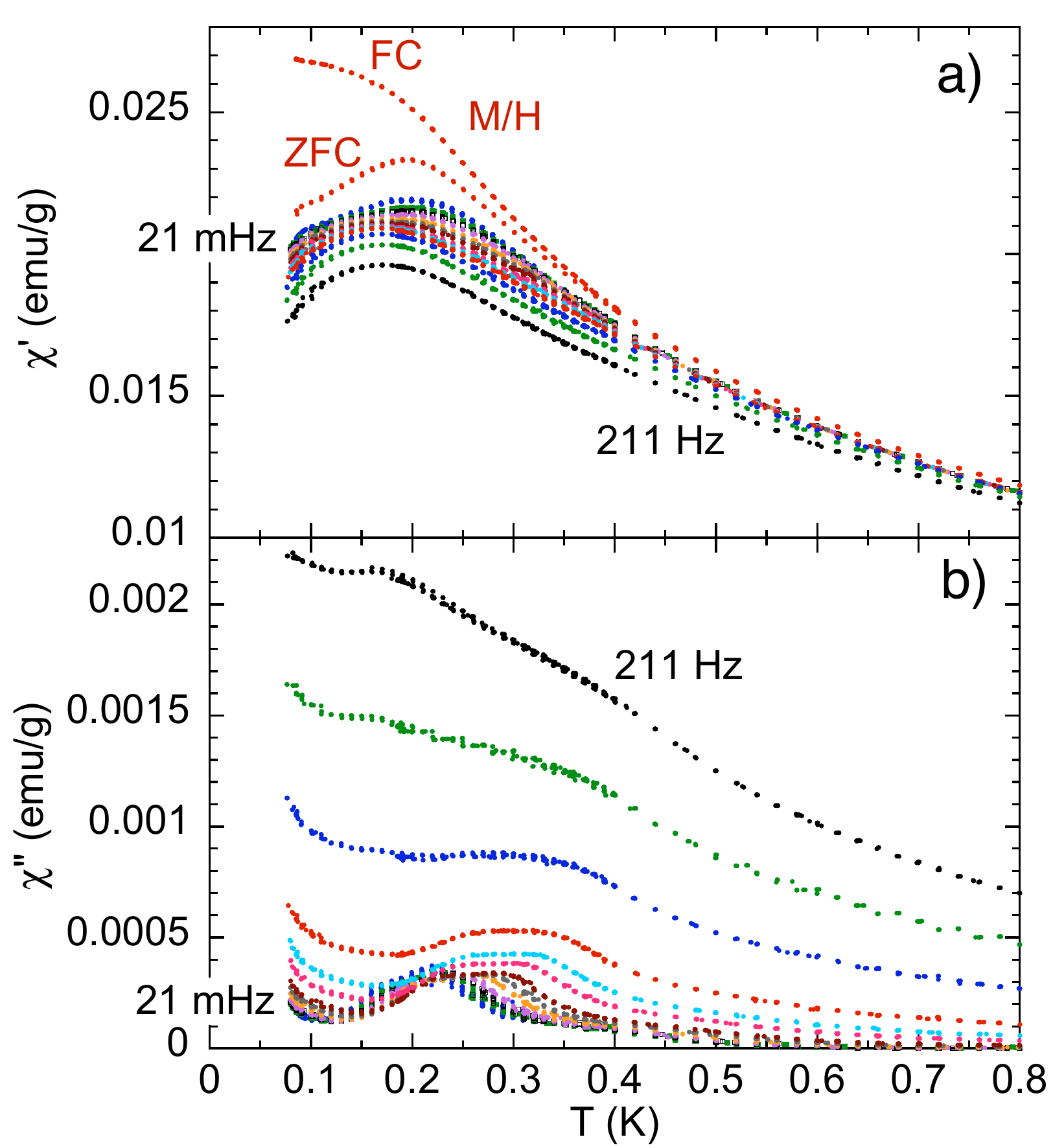}
\caption{(Color online) Susceptibility measurements as a function of temperature along [111] in sample B. a) M/H measured in a ZFC-FC procedure with $H_{\rm DC}=5$ Oe together with the real part $\chi'$ of the AC susceptibility. b) Imaginary part $\chi''$ of the AC susceptibility, for frequencies between 21 mHz and 211 Hz, with $H_{\rm AC}=0.4$ Oe.}
\label{figXTB}
\end{figure}

The Zero Field Cooled - Field Cooled (ZFC-FC) curve of the DC susceptibility measured in sample B in a small field ($H_{\rm DC}=5$ Oe) presents an irreversibility below 250 mK, in a very similar way as samples C1 along [111] and C2 along [110] (See Figure \ref{figXTB}(a)). Associated with this irreversibility, the real part $\chi'$ of the AC susceptibility presents a broad peak which depends weakly on the frequency, as shown in  Figure \ref{figXTB}(a). The imaginary part $\chi''$ shows a more complex behavior (See Figure \ref{figXTB}(b)). The fact that the $\chi_{\rm AC}$ peaks are better defined in samples C1 and C2 may indicate that the distribution of relaxation times is narrower in these samples than in sample B. The presence of local inhomogeneities, by modifying the local energy environment of the spins might extend such a distribution, resulting in broader features in AC susceptibility measurements.
Interestingly, for frequencies larger than 21 Hz, $\chi''$ presents the same tail as observed in samples C1 and C2 when the temperature is increased, falling to zero at much higher temperatures. 

\begin{thebibliography}{}
\bibitem{footnote1} A better way to determine the freezing temperature is to find the temperature at which the $\chi'$ curve leaves the equilibrium susceptibility for each frequency. This would be very difficult to do with our $\chi'$ data, and further reinforces the idea that the behavior is not of a canonical spin-glass.
\end{thebibliography}
\end{document}